\documentclass[useAMS,usenatbib]{mn2e}

\usepackage{graphicx}
\usepackage{hyperref}
\usepackage{multirow}

\newcommand{\be}{\begin{equation}}
\newcommand{\ee}{\end{equation}}
\newcommand{\bea}{\begin{eqnarray}}
\newcommand{\eea}{\end{eqnarray}}

\newcommand{\mm}{\,\hbox{mm}}
\newcommand{\km}{\,\hbox{km}}
\newcommand{\mum}{\,\mu\hbox{m}}
\newcommand{\cm}{\,\hbox{cm}}

\newcommand{\AU}{\,\hbox{AU}}
\newcommand{\g}{\,\hbox{g}}

\newcommand{\Gyr}{\,\hbox{Gyr}}

\newcommand{\K}{\,\hbox{K}}

\title[Dust in systems with transiting planets]
{Debris disc candidates in systems with transiting planets}

\author[A.V. Krivov et al.]
       {
       A. V. Krivov\thanks{krivov@astro.uni-jena.de (AVK)},
       M. Reidemeister, S. Fiedler, T. L\"ohne, and R. Neuh\"auser\\
       Astrophysikalisches Institut, Friedrich-Schiller-Universit\"at Jena,
       Schillerg\"a{\ss}chen~2--3, 07745 Jena, Germany
       }

\begin{document}

\date{Accepted {\em 2011 August 15}. Received {\em 2011 August 11};
in original form {\em 2011 June 1}}

\pagerange{L\pageref{firstpage}--L\pageref{lastpage}} \pubyear{2011}

\maketitle

\label{firstpage}

\begin{abstract}
Debris discs are known to exist around many planet-host stars, but no debris dust has 
been found so far in systems with transiting planets.
Using publicly available catalogues,
we searched for infrared excesses in such systems.
In the recently published Wide-Field Infrared Survey Explorer ({\it WISE})
catalogue, we found 52 stars with transiting planets.
Two systems with one transiting ``hot Jupiter'' each, TrES-2 and XO-5,
exhibit small excesses both at $12\mum$ and $22\mum$
at a $\ga 3 \sigma$ level.
Provided that one or both of these detections are real,
the frequency of warm excesses in systems with transiting planets of
2-4\% is comparable to
that around solar-type stars probed at similar wavelengths with {\it Spitzer}'s
MIPS and IRS instruments.
Modelling suggests that the observed excesses would stem from dust rings with radii
of several $\AU$.
The inferred amount of dust is close to the maximum expected theoretically
from a collisional cascade in asteroid belt analogues.
If confirmed, the presence of debris discs in systems with transiting planets
may put important constraints onto formation and migration scenarios of hot Jupiters.
\end{abstract}

\begin{keywords}
planetary systems --
planets and satellites: detection --
circumstellar matter --
stars: individual: XO-5, HAT-P-5, TrES-2 parent star, CoRoT-8
\end{keywords}


\section{Introduction}
Many debris discs have been found in systems with known radial velocity (RV) planets
\citep[e.g.][]{beichman-et-al-2005a,moromartin-et-al-2007,%
trilling-et-al-2008,bryden-et-al-2009,kospal-et-al-2009},
and a few systems with debris discs and directly imaged planetary candidates
are known
\citep{kalas-et-al-2008,marois-et-al-2008,lagrange-et-al-2010}.
However, debris dust has not been found yet
in systems with planets detected by transits.

\looseness=-1
In this paper, we search for debris dust in systems with transiting
planets, using publicly available catalogues of transiting planets and several infrared (IR)
surveys.
The motivation is obvious.
A successful search would extend the list of known ``full'' planetary systems that 
harbour both
planets and asteroid or Kuiper belt analogs.
Furthermore, it is the transit technique that allows 
determination of many 
planetary parameters, such as masses, radii and densities, and can provide insights 
into properties of  planetary atmospheres and interiors.
Finally, transiting planets are on the average even closer
to their parent stars than those discovered by the RV method, which might
be related to somewhat different formation circumstances.
Therefore, systems with transiting planets are of special interest.
Detection of planetesimal belts, which are leftovers of planet formation, could help 
constraining various formation and evolution scenarios of those planets.
And conversely, precise knowledge
of planetary parameters could put constraints on the properties of the planetesimal belts.
For instance, accurate masses and orbits of planets would result in tighter constraints
on dynamical stability zones and thus location of planetesimal belts.


\section{Search for dust}

\label{sec:search}

A list of 93 currently known
systems with transiting planets was taken 
from \url{exoplanets.org} \citep{wright-et-al-2011}\footnote{Accessed on May 11, 2011.}.
This list was compared with target lists of several IR missions:
{\it IRAS} \citep{neugebauer-et-al-1984},
{\it ISO} \citep{kessler-et-al-1996},
{\it Spitzer} \citep{werner-et-al-2004},
{\it AKARI} \citep{murakami-et-al-2007},
and {\it WISE} \citep{wright-et-al-2010},
which we accessed through IRSA, the NASA/IPAC 
Infrared Science Archive at {\url{http://irsa.ipac.caltech.edu}.

\subsection{{\it IRAS}, {\it ISO}, {\it Spitzer}, and {\it AKARI}}

Nearly  all of the transit planet host stars are located at hundreds of parsecs
from the Sun and are thus faint.
Accordingly, we had not expected to find them in older, and shallower, {\it IRAS} and {\it ISO}
catalogues and indeed, have not found any.
For example, of 93 systems with transiting planets listed in \url{exoplanets.org},
only five are within 50~pc.
These are GJ 436, GJ 1214, HAT-P-11, HD 189733, and  HD~209458.
None of them appears in {\it IRAS}, {\it ISO}, and {\it WISE} catalogues.
Three of them, GJ~436, HD~189733, and HD~209458,
have been probed by {\it Spitzer}/MIPS at $24$ and $70\mum$, yielding no excess detection 
\citep{bryden-et-al-2009}.
We found an entry for the latter star in the {\it AKARI} catalogue,
reporting a detection at $9\mum$, which is consistent with the photospheric level.
Note that HD~209458b
was the first exoplanet found to transit the disc of its parent star 
\citep{charbonneau-et-al-2000}.

We have also identified two more distant transit planet host stars
that were observed by {\it Spitzer}/MIPS:
HD~80606 and HD~149026.
No excess at $24$ and $70\mum$ was found for
HD 189733 \citep{bryden-et-al-2009}.
For HD~80606,
the result is ambiguous due to pointing problems \citep{carpenter-et-al-2008}.
{\it AKARI} has observed one more transit host star, too:
HD~149026. It has been detected at $9\mum$, showing no excess.

\subsection{\it WISE}

The search in the {\it WISE} Preliminary Source Catalog was more successful.
This is perhaps not a surprise,
given the broad sky coverage (57\%) of the catalogue,
an excellent sensitivity of the instrument
and thus a huge number of sources observed (257 million).
The {\it WISE} catalogue provides measured magnitudes in four bands
$W_i$ ($i = 1$, \ldots, $4$), which are centred at $3.4$, $4.6$, $12$, and $22\mum$.
Of 93 systems with transiting planet candidates listed in \url{exoplanets.org},
we found 53 with entries in the {\it WISE} catalogue.
One source~--- CoRoT-14~--- is irretrievably contaminated by
ghost images in bands $W_3$ and $W_4$ and was excluded from further
analysis.

To select possible IR excess candidates amongst the remaining 52 sources,
we first converted the observed magnitudes
in the four bands into spectral flux densities.
Since no excesses are expected in bands $W_1$ and $W_2$,
we made simple photospheric predictions for $W_3$ and $W_4$ 
from the $W_1$ and $W_2$ fluxes.
At first, we roughly corrected the $W_1$ and $W_2$ fluxes
for an expected average level of interstellar extinction.
Considering that systems in our sample are typically at a few hundreds parsecs,
we set $A_V$ to $0.5^m$,
which translates to $A(W_1) = 0.029^m$ and $A(W_2) = 0.012^m$ 
\citep{rieke-lebofsky-1985}.
We then fitted the corrected $W_1$ and $W_2$ fluxes with a power law
$F_{\rm phot} = F_{\rm phot}^0 \lambda^{-b}$, 
with $F_{\rm phot}^0$ and $b$ being the fitting parameters.
Subtracting the expected photospheric flux from the observed one,
we derived the ``excess flux'' $F \equiv F_{\rm obs}-F_{\rm phot}$
in bands $W_3$ and $W_4$.
The net uncertainty of a photometric point for a given star in the band $W_3$
or $W_4$ was computed as
$\sigma = \sqrt{\sigma_{\rm phot}^2+ \sigma_{\rm obs}^2+ \sigma_{\rm cal}^2}$.
Here, $\sigma_{\rm phot}$ is the photospheric uncertainty, which we estimated
from the combined uncertainties of the measurements in bands $W_1$ and $W_2$,
given in the {\it WISE} catalogue.
Next, $\sigma_{\rm obs}$ is the measurement uncertainty in the bands of interest,
$W_3$ and $W_4$, also taken from the {\it WISE} catalogue.
Finally,
$\sigma_{\rm cal}$ is the absolute calibration uncertainty of the
{\it WISE} instrument ($2.4$, $2.8$, $4.5$, and $5.7$\% for bands from $W_1$ to $W_4$).
The significance of an excess can now be defined as $\chi = F /\sigma$.

The distributions of $\chi$-values in the sample are shown in Fig.~\ref{fig:wise}
for bands $W_3$ (top) and $W_4$ (bottom).
The $W_3$ histogram appears close to gaussian, without
any obvious outliers.
However, the $W_4$ histogram uncovers a bin containing four systems with 
$\chi > 1.75\sigma$, clearly separated from the gaussian bulk.
These are XO-5, HAT-P-5, TrES-2, and CoRoT-8.
These four excess candidates will be checked in  
Sect.~\ref{sec:excesses} more thoroughly,
including an in-depth photospheric analysis and more accurate uncertainty estimates.

\begin{figure}
  \begin{center}
  \includegraphics[width=0.35\textwidth]{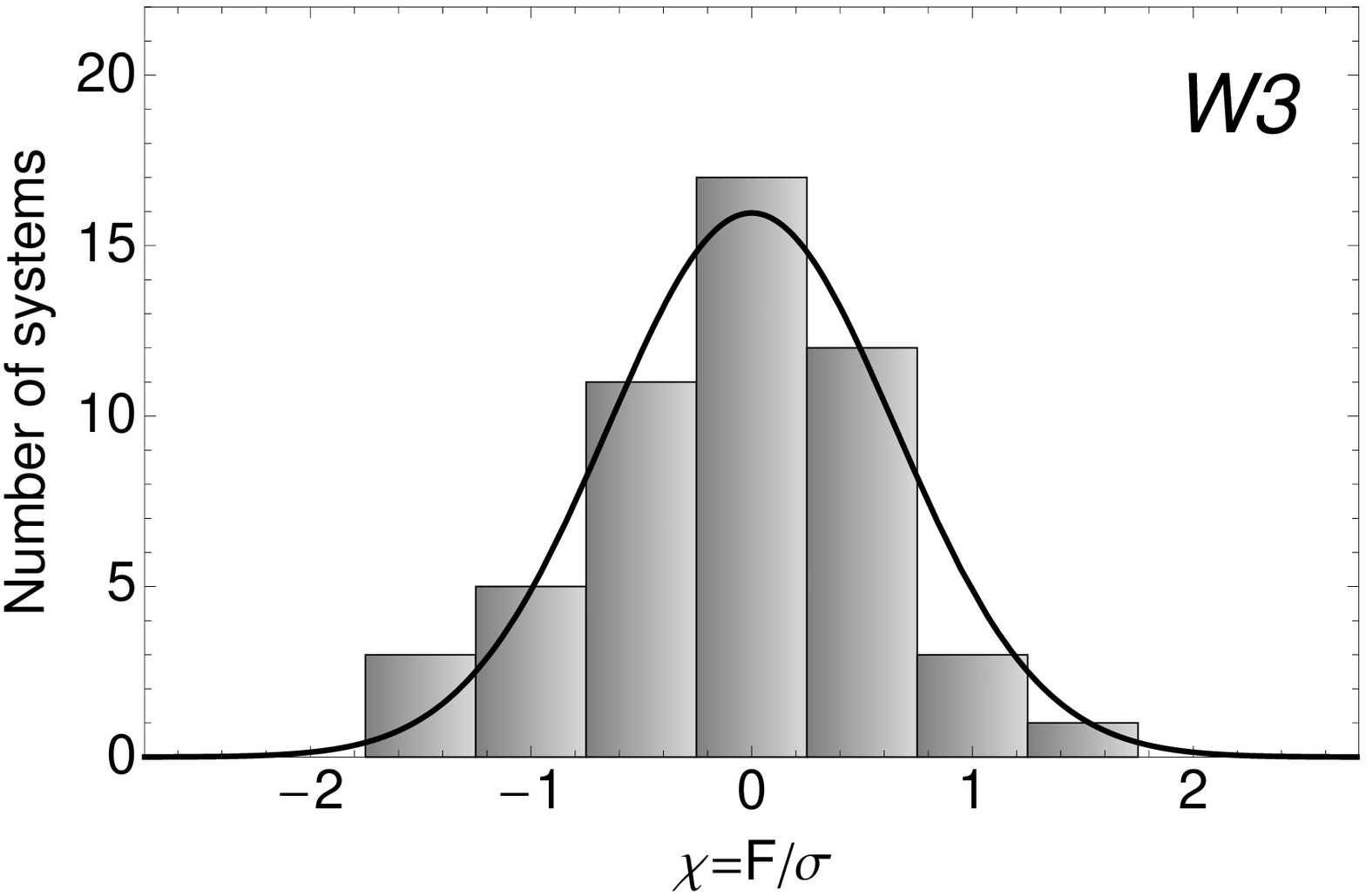}\\
  \includegraphics[width=0.35\textwidth]{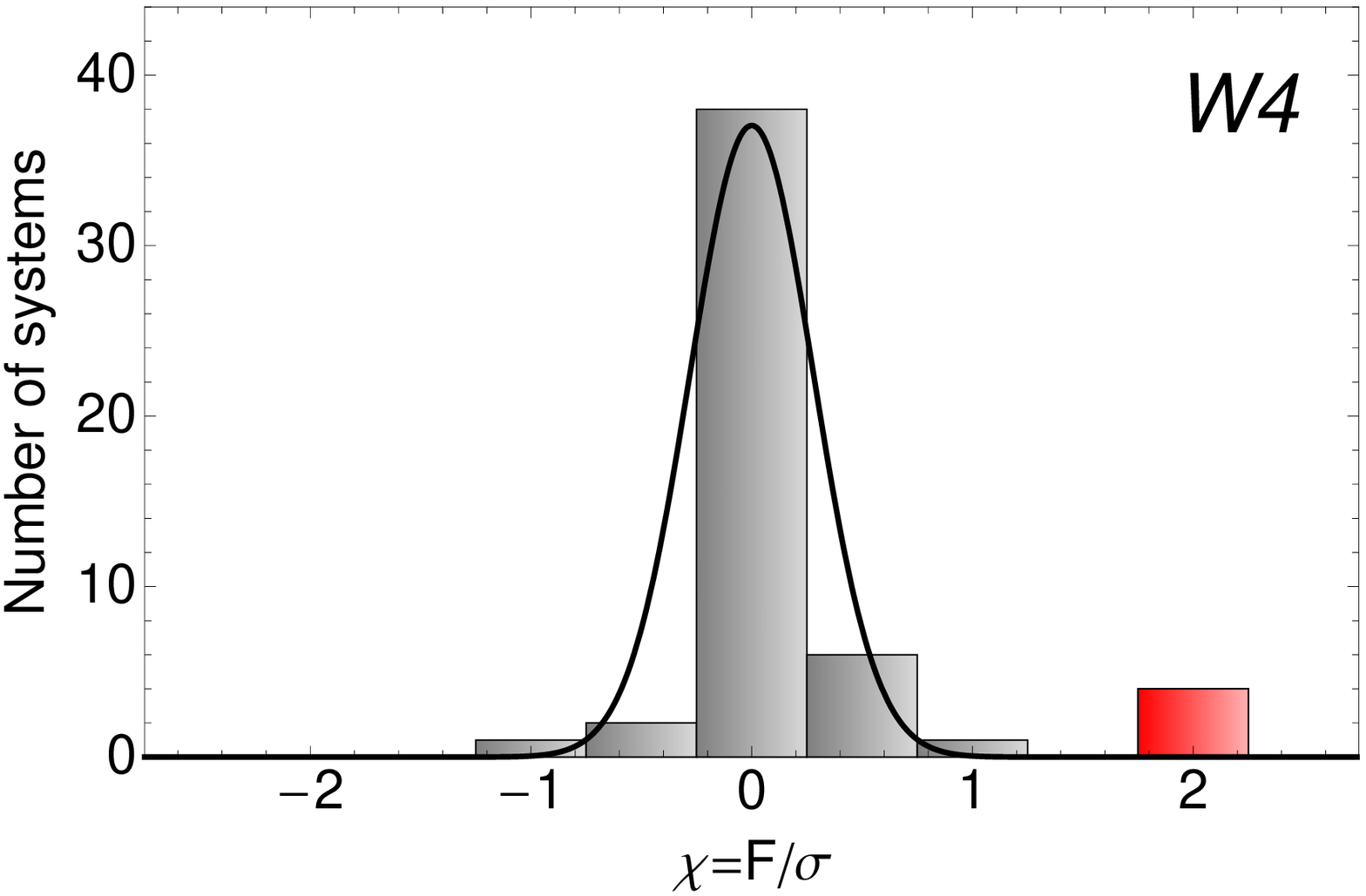}
  \end{center}
  \caption{
  Histogram of $\chi = F/\sigma$ values at $12\mum$ (top)
  and $22\mum$ (bottom) for systems with transiting planets.
  For comparison, overplotted are closest gaussian distributions, which
  have variances of 0.65 (top) and 0.28 (bottom).
  That these variances are smaller than unity may suggest that the procedure
  of calculating $\sigma$,  described in the text, is too cautious and
  overestimates the actual noise.
  The rightmost bin in the $W_4$ histogram shown in red contains four excess
  candidates.
  We checked that slight modifications to the photospheric extrapolation procedure
  or extinction correction (i.e. using $A_V$ values between $0.0^m$ and $1.0^m$)
  slightly skew and shift the W3 histogram, but preserve the
  W4 one, including the bin with the four outliers.
  }
  \label{fig:wise}
\end{figure}


\section{Analysis of excess candidates}
\label{sec:excesses}

For the excess candidates, we have collected
stellar data (Tab.~\ref{tab:stars}) as well as
optical and near-IR photometry.
In the visual, we used the USNO-B1.0 Catalog \citep{monet-et-al-2003},
the Guide Star Catalog, Vers.~2.3.2 \citep{lasker-et-al-2008},
and the All-Sky Compiled Catalogue of 2.5 Million Stars \citep{kharchenko-roeser-2009}.
The near-IR data stem from the 2MASS All-Sky Catalog of Point Sources
\citep{skrutskie-et-al-2006}.
For transforming the $B$, $V$, $R$, $I$ magnitudes into units of flux density [Jy], we used
the Johnson calibration system and for the 2MASS $J$, $H$,
$K_s$ bands the \citet{cohen-et-al-2003} calibrations.

\begin{table*}
  \caption{
  Stellar parameters.
  \label{tab:stars}
  }
  \begin{tabular}{lllccllcll}
   \hline
   Star		&V$_{\rm mag}$ & $d$ [pc]	& SpT    & $M_\star$ [$M_\odot$]& $T_{\rm eff}$[$\K$]& $R_\star$  [$R_\odot$]	& Ref.	&$d$ [pc]	& $T_{\rm eff}$[$\K$]\\
   \hline
   XO-5		& 12.1	   &  $260 \pm 12$	& late G & $0.88 \pm 0.03$	& $5370 \pm 70$	     & $1.08 \pm 0.04$		& [1] 	&$286$ ($+10$\%) & $5800$ ($+8$\%)\\  
   HAT-P-5	& 12.0	   &  $340 \pm 30$	& early G& $1.16 \pm 0.06$	& $5960 \pm 100$     & $1.17 \pm 0.05$		& [2] 	&$340$ ($\pm0$\%)& $6400$ ($+7$\%)\\
   TrES-2	& 11.4	   &  $230$		& G0~V	 & $1.08 \pm 0.11$	& $5960 \pm 100$     & $1.00^{+0.06}_{-0.04}$	& [3]	&$219$ ($-5$\%)  & $6400  (+7$\%) \\
   CoRoT-8	& 14.8	   &  $380 \pm 30$	& K1~V	 & $0.88 \pm 0.04$	& $5080 \pm 80$	     & $0.77 \pm 0.02$		& [4]	&$304$ ($-20$\%) & $5400$ ($+6$\%)\\
   \hline
   \end{tabular}

\smallskip
{\small
{\em References:}
[1]~\citet{pal-et-al-2009},
[2]~\citet{bakos-et-al-2007},
[3]~\citet{odonovan-et-al-2006},
[4]~\citet{borde-et-al-2010}.\\
{\em Note:} The last two columns list best-fit values and their deviation from the starting values.
}
\end{table*}

\begin{figure}
  \begin{center}
  \includegraphics[width=0.36\textwidth]{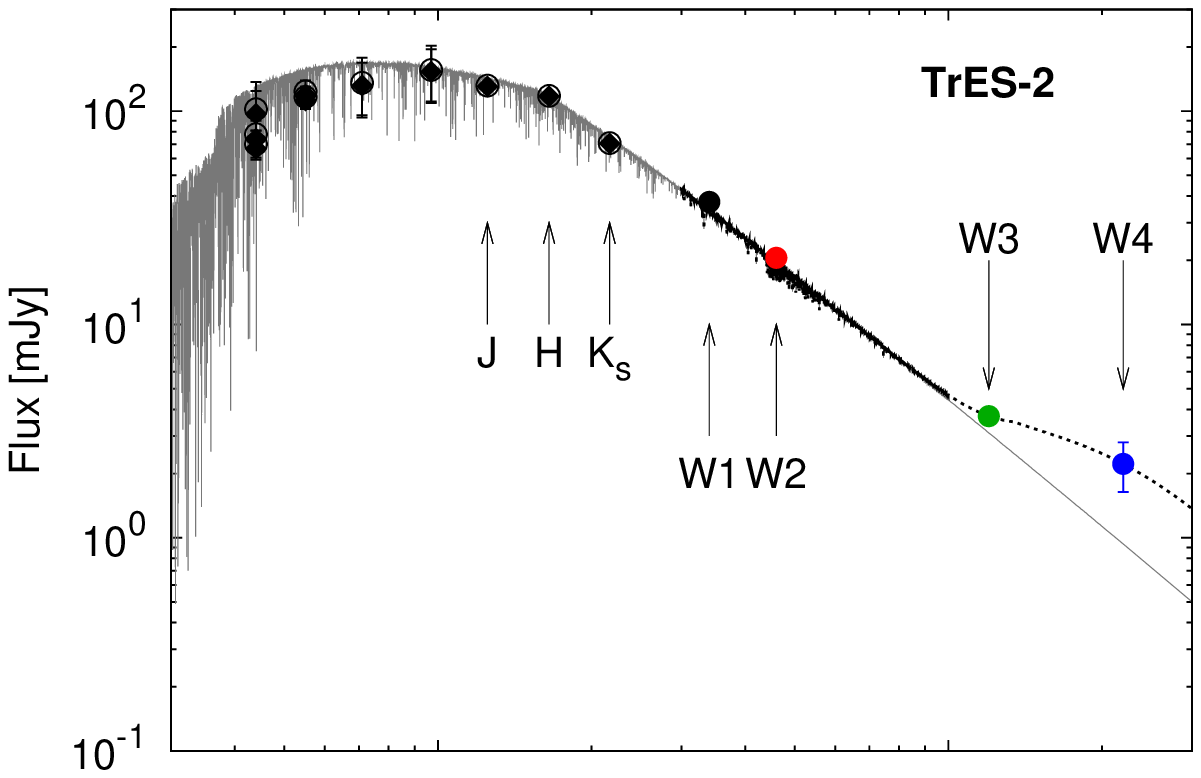}\\[-5mm]
  \includegraphics[width=0.36\textwidth]{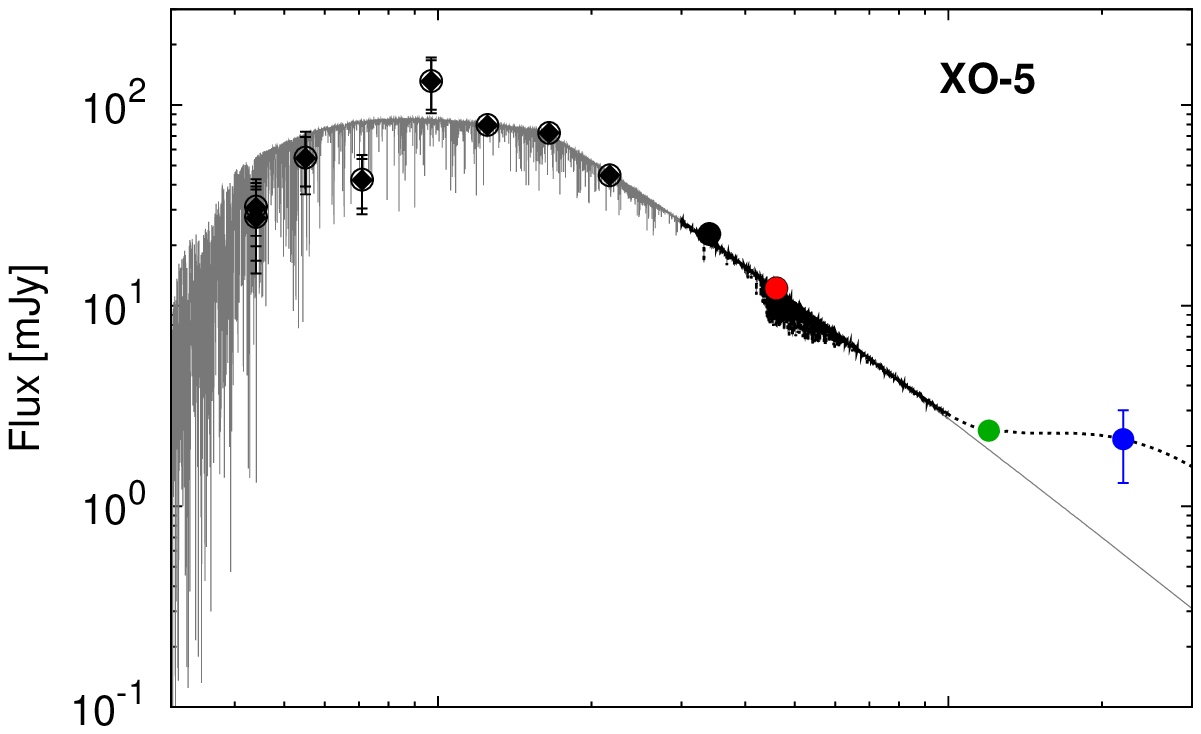}\\[-5mm]
  \includegraphics[width=0.36\textwidth]{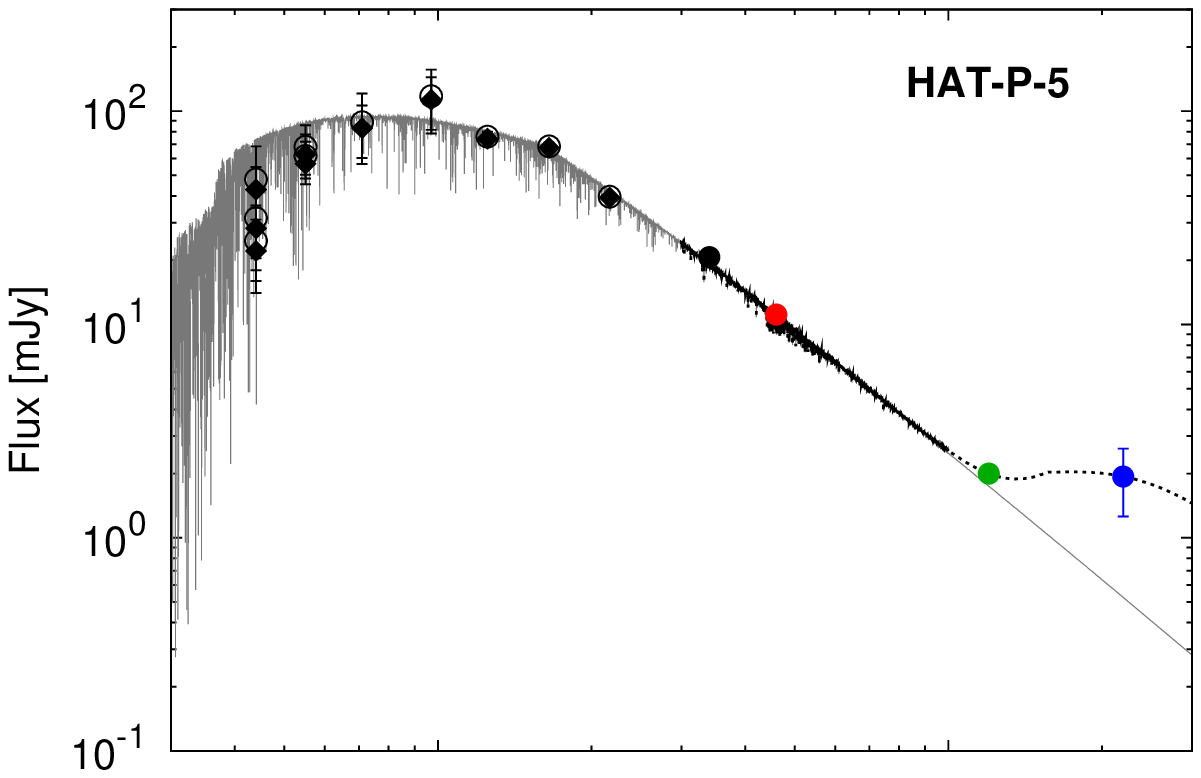}\\[-5mm]
  \includegraphics[width=0.36\textwidth]{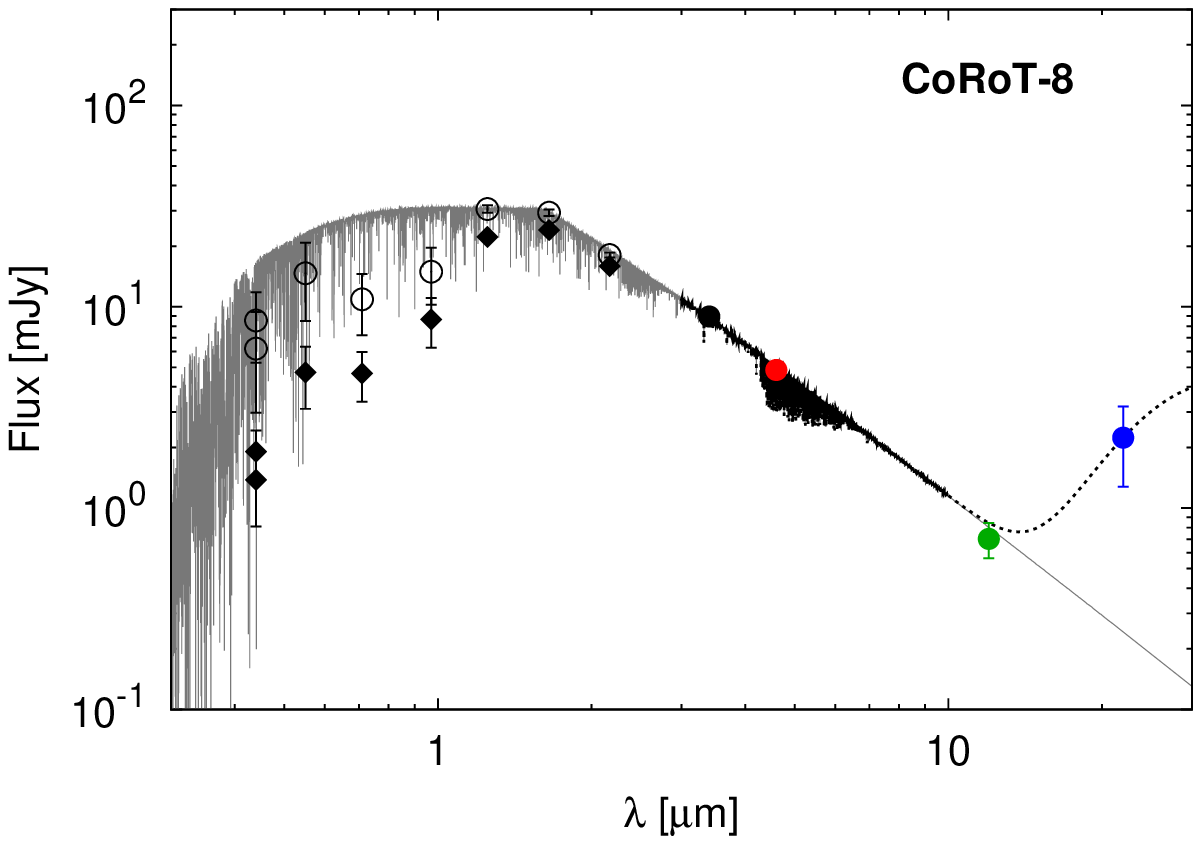}
  \end{center}
  \caption{
  Spectral energy distributions (SEDs) of four selected stars.
  Grey solid line: predicted extinction-corrected photosphere.
  Diamonds and open circles: visual and near-IR photometry data 
  before and after correction for interstellar extinction, respectively.
  Filled circles: extinction-corrected {\it WISE} data (errors bars are $\sigma_{\rm obs}$).
  Black dashed line: modified BB model.
  }
  \label{fig:sed}
\end{figure}

At the first step, this photometry was corrected for interstellar extinction.
Since for distances considered here the latter is known to correlate with distance only
weakly, we used colour indices from \citet{kenyon-hartmann-1995} and the spectral type
as given in Tab.~\ref{tab:stars} to derive the best-fit $A_V$ from multiple colours
and then $A_\lambda/A_V$ ratios of \citet{rieke-lebofsky-1985} to compute
extinction for the wavelengths of all photometry points.
The derived $A_V$ values are within $0.1^m$ for non-CoRoT stars, but as large as
$1.3^m\pm 0.2^m$ for CoRoT-8.
The extinction in W3 and W4 bands does not exceed
$0.04^m$ (CoRoT-8).

\looseness=-1
At the second step,
we performed a minimum $\chi^2$ fitting of the extinction-corrected
stellar photospheric fluxes by NextGen models \citep{hauschildt-et-al-1999},
only to data points between $1\mum$ and  $5\mum$.
This is because those wavelengths are short enough not to expect
any excess emission, but are long enough for interstellar extinction
to be small.
The interval from $1\mum$ to $5\mum$ includes three 2MASS points
($J$, $H$, $K_s$) and two {\it WISE} points ($3.4$ and $4.6\mum$),
all of which were given equal weights.
Since the surface gravity of our stars ($\log g$ between $4.33$ and $4.61$)
and their metallicity ([Fe/H] from $-0.2$ to $+0.31$)
deviate from the solar values only slightly
(see Tab.~\ref{tab:stars} for references),
we assumed a log $g$ of 4.5 and the solar metallicity.
Using $T_{\rm eff}$ and $d$ listed in Tab.~\ref{tab:stars} as starting
values, we varied the temperature by $\pm 400\K$ in $200\K$ steps
and the distance to the star within $\pm 30\%$ in 5\% steps
to derive best-fit values of these two parameters.
This yielded deviations from the starting values
of up to 8\% in $T_{\rm eff}$ and up to 20\% in $d$ (Tab.~\ref{tab:stars}).
The results with the photometric points overplotted are shown in Fig.~\ref{fig:sed}.
Importantly, TrES-2 and XO-5 and possibly also HAT-P-5 reveal
small excesses in band W3 as well, which were not seen in Fig.~\ref{fig:wise} (top).

\begin{table*}
  \caption{
  Fluxes [mJy], uncertainties [mJy], and significance of excesses.
  \label{tab:excesses}
  }
  \begin{tabular}{lccccccccccc}
   \hline
System           &  Band  & $F^\star_{\rm obs}$ &$F_{\rm obs}$& $F_{\rm phot}$ & $F$   & $\sigma_{\rm phot}$ & $\sigma_{\rm obs}$ & $\sigma_{\rm cal}$ & $\sigma$ & $\chi$ & $\chi_{\rm joint}$\\
   \hline
TrES-2           &  W3    &  3.71               & 3.72        &  3.10          &  0.62 &  0.20               &  0.11              &  0.17              &  0.29    &  2.17  & \multirow{2}{*}{3.28}\\
                 &  W4    &  2.22               & 2.22        &  0.93          &  1.29 &  0.12               &  0.58              &  0.13              &  0.61    &  2.12  & \\
XO-5             &  W3    &  2.38               & 2.38        &  1.92          &  0.47 &  0.09               &  0.14              &  0.11              &  0.20    &  2.35  & \multirow{2}{*}{3.23}\\
                 &  W4    &  2.16               & 2.16        &  0.58          &  1.58 &  0.08               &  0.85              &  0.12              &  0.86    &  1.84  & \\
HAT-P-5          &  W3    &  1.99               & 2.00        &  1.74          &  0.26 &  0.09               &  0.10              &  0.09              &  0.16    &  1.58  & \multirow{2}{*}{2.82}\\
                 &  W4    &  1.93               & 1.94        &  0.53          &  1.41 &  0.09               &  0.68              &  0.11              &  0.70    &  2.03  & \\
CoRoT-8          &  W4    &  2.17               & 2.24        &  0.24          &  2.00 &  0.09               &  0.96              &  0.13              &  0.97    &  2.05  & \\
   \hline
   \end{tabular}

\smallskip
{\small
{\em Columns:} Observed flux $F^\star_{\rm obs}$,
observed flux after correction for extinction $F_{\rm obs}$,
expected photospheric flux $F_{\rm phot}$,
excess flux $F$;
uncertainty of photospheric flux $\sigma_{\rm phot}$,
observation uncertainty $\sigma_{\rm obs}$,
absolute calibration uncertainty $\sigma_{\rm cal}$,
net uncertainty of the excess flux $\sigma$,
excess significance level in a single band $\chi$;
joint ($W_3$ and $W_4$) significance level $\chi_{\rm joint}$.
}
\end{table*}

\looseness=-1
We now come to a detailed analysis of the fluxes and their uncertainties.
Denote the observed flux by $F^\star_{\rm obs}$,
the extinction-corrected one by $F_{\rm obs}$,
the predicted extinction-corrected photospheric flux by $F_{\rm phot}$,
and the excess flux by $F \equiv F_{\rm obs}-F_{\rm phot}$.
As in Sect.~\ref{sec:search}, the net uncertainty of $F$
for a given star in band $W_3$ or $W_4$ is computed as
$\sigma = \sqrt{\sigma_{\rm phot}^2+ \sigma_{\rm obs}^2+ \sigma_{\rm cal}^2}$.
The measurement uncertainty $\sigma_{\rm obs}$ and
the calibration uncertainty $\sigma_{\rm cal}$ are included
as described before.
However, the photospheric uncertainty  $\sigma_{\rm phot}$ is 
now a by-product of the fitting procedure.
It is dominated by a scatter in $J$, $H$, $K_s$, $W_1$, and $W_2$ points
(the error bars of the points themselves as well as the uncertainty of the
extinction correction are much smaller).
All the quantities above, and the resulting
excess significance $\chi = F /\sigma$,
are listed in Tab.~\ref{tab:excesses}.
Nearly all excesses are at $\approx 2\sigma$ level, whereas
usually a $3\sigma$ excess is treated as a significant detection.
However, in the cases of TrES-2, XO-5, and HAT-P-5, the excess is detected
in two bands.
The combined multi-band ($W_3$ and $W_4$) gaussian statistics suggests
the significance level for these sources of
$3.28$, $3.23$, and $2.82$, respectively.
This finally selects two systems, TrES-2 and XO-5, as $> 3\sigma$-significant
and thus the best excess candidates.
The binomial probability that one of these two detections
is false is only $6.4$\%, and
the probability that both are false is as low as $0.2$\%.
The expected number of false detections at $>3\sigma$ level is just 0.14;
we detected two excesses at that level.

\begin{table*}
  \caption{
  Dust parameters inferred from the observed excesses and parameters of transiting planets.
  \label{tab:results}
  }
  \centering
  \begin{tabular}{l|rl|ccrccc|cll}
   \hline
 System & \multicolumn{2}{|c}{Blackbody} & \multicolumn{6}{|c|}{Modified blackbody}							& \multicolumn{3}{c}{Planet}     \\
   \cline{4-9}
        & $T_{\rm d}$ & $r_{\rm d}$ & $s_{\rm blow}$ & $s_0$	& $T_{\rm d}$     & $r_{\rm d}$ & $M_{\rm d}$       & $f_{\rm d}$       & $a_{\rm pl}$ & $e_{\rm pl}$ & $M_{\rm pl}$   \\
        & [$\K$]      & [$\AU$]	    & [$\mum$]       & [$\mum$]	& [$\K$]	  & [$\AU$]     & [$M_\oplus$]      &                   & [$\AU$]      &              & [$M_{\rm jup}$]\\
   \hline
TrES-2 	&  218        & 1.7	    & 0.4	     & 2.1	& 155   	  & 5.8         & $5\times 10^{-5}$ & $3\times 10^{-4}$	& 0.037        & 0 (fixed)& 1.28       \\
XO-5  	&  181        & 2.2	    & 0.4	     & 2.0	& 133   	  & 8.0         & $1\times 10^{-4}$ & $6\times 10^{-4}$	& 0.051        & 0.049    & 1.06       \\
   \hline
   \end{tabular}

\smallskip
{\small
{\em Note:} Planetary parameters are taken from the papers listed in Tab.~\ref{tab:stars}.
}
\end{table*}


\section{Presumed dust belts}
\label{sec:parms}

In what follows, we estimate the parameters of dust that would produce the excesses
in the best candidate systems, TrES-2 and XO-5, provided these are real.
Since the excesses are of low significance and the data are limited to two 
photometry points,
a detailed SED modelling based on various assumptions about the size distribution
and composition of dust is not warranted.
Instead, we used
a pure blackbody (BB) and a modified BB emission model.
In the latter case, we assumed a single grain size $s_0$
and the opacity index of $-2$ beyond $\lambda = 2\pi s_0$.
The effective grain size $s_0$ was chosen in the following way.
Assuming a power-law size distribution with the index $q=3.5$
and the lower cutoff radius $s_{\rm min}$ of twice the radiation pressure
blowout limit $s_{\rm blow}$ \citep[see. e.g.,][]{krivov-et-al-2006,thebault-augereau-2007},
we have equated the emission of a disc of grains having such a size distribution
and the emission of a disc composed of equal-sized grains of radius~$s_0$:
\bea
  &&
  \int_{s_{\rm min}}^\infty
  Q_{\rm abs}(\lambda, s) \;
  B_\nu (\lambda, T_{\rm d}(r_{\rm d}, s)) \;
  s^{2-q} \; ds
\nonumber\\
  &=&
  Q_{\rm abs}(\lambda, s_0) \;
  B_\nu (\lambda, T_{\rm d}(r_{\rm d}, s_0)) \;
  s_0^2 \;
  \int_{s_{\rm min}}^\infty
  s^{-q} \; ds ,
\label{size eq}
\eea
where
$r_{\rm d}$ is the distance from the star,
$\lambda$ is the wavelength where excess emission is observed,
$B_\nu$ is the Planck intensity,
$Q_{\rm abs}(\lambda, s)$ is the grain absorption efficiency,
and $T_{\rm d}$ is the grain temperature.
In calculating $s_{\rm blow}$, we assumed the unit radiation pressure efficiency
and the bulk density of $3\g\cm^{-3}$ and took the stellar parameters
from Tab.~\ref{tab:stars}.
Equation~(\ref{size eq}) was solved for $s_0$.

We then sought pure BB and modified BB curves that reproduce
$F(12\mum)$ and $F(22\mum)$.
This has yielded estimates of
the temperature, location, mass,
and the fractional luminosity of the emitting dust
(Tab.~\ref{tab:results}).
When deriving the dust mass, we converted the mass of grains with $s=s_0$
into the mass of grains with $s<1\mm$,
assuming a power-law size distribution with a slope of $3.5$.
Both stars, TrES-2 and XO-5, appear to have rings with radii of 6--8$\AU$
and fractional luminosities
in the range (3--6)$\times 10^{-4}$.
We stress that all these inferred values
are quite uncertain, because they rest on
scarce photometric data and their derivation involves a number of simplifying
assumptions and poorly known parameters.


\section{Conclusions and discussion}

We have found that
two out of 52 systems with transiting planets observed by {\it WISE} 
reveal warm two-band ($12$ and $22\mum$) IR excesses at $> 3\sigma$ level. 
The probability that both excesses are real is 94\% and that
one of them is real is $99.8\%$.

Provided that one of the two systems, or both, do possess a warm disc,
this would imply the excess incidence rate of $2$--$4$\%.
For comparison, the {\it Spitzer}/MIPS detection rate of $24\mum$ excesses
around old ($\sim 4\Gyr$) field stars was found to be  1/69 ($1 \pm 3$\%)
\citep{bryden-et-al-2006}.
Another sample of solar-type stars probed by MIPS at $24\mum$
resulted in $\approx 4$\% detection rate, averaged over all ages
\citep{trilling-et-al-2008}.
\citet{lawler-et-al-2009} analyzed {\it Spitzer}/IRS observations of nearby solar-type
stars and found excess around $12$\% of them
in the long-wavelength IRS band ($30$-$34\mum$),
but only $1$\% of the stars have detectable excess
in the short wavelength band ($8.5$-$12\mum$).
Thus the frequency of warm excesses around solar-type stars
with transiting planets seen in the {\it WISE} data
may be comparable to that in
unbiased samples of similar stars found with {\it Spitzer}.

Each of the two systems discussed here hosts one known close-in planet
and, if the excesses are real, an asteroid belt-size dust ring well outside
the planetary orbit.
In both cases,
more planets could orbit both inside and outside the
belts.
Additional planets at $\la 10\AU$ could be revealed by in-depth
RV analyses, by transits,
or by transit time variations of already known planets
\citep{maciejewski-et-al-2011}.
The latter method was used for
TrES-2 \citep{raetz-et-al-2009} and XO-5 \citep{maciejewski-et-al-2011b}.
Non-detection is
consistent with the presence of debris belts at several AU,
which are incompatible with planets in that region.
In the case of TrES-2, \citet{raetz-et-al-2009}
noticed a second dip in the light curve,
both in their own
light curves and those published in the literature.
This second dip has been observed several times and then disappeared.
In addition to other possible reasons for this
effect, discussed by \citeauthor{raetz-et-al-2009}, it could be
due to an occultation by material in the debris disc.
Estimates show that a clump of dust produced
in a recent collision of two $\sim 100\km$-sized planetesimals
would bear enough cross section to account for such a dip
before it is azimuthally spread into a ring in a few years,
although the probability of witnessing such an event is low.

Planets at largest orbital radii ($\ga 10\AU$) will be hard to find
by the transit technique.
Direct imaging and astrometry are not feasible either, since these systems are too old
and too distant.
It will also be difficult to search for possible
Kuiper belt analogues on the periphery of the systems,
because they are too faint for far-IR facilities such as {\it Herschel}.
However, future mid-IR instruments such as {\it JWST/MIRI} should have enough sensitivity
to study warm dust in great detail, including dust grain spectroscopy.
They may also take a closer look at further potential 
excess candidates such as HAT-P-5 and CoRoT-8 identified in this study.

The origin and the production mechanisms of the presumed dust are unclear.
We have computed the dust mass expected to be produced through
a steady-state collisional cascade in a belt of ``asteroids''
with moderate eccentricities,
using the model of \citet{loehne-et-al-2007} with a velocity-dependent
critical fragmentation energy from \citet{stewart-leinhardt-2009}.
At ages of $\sim 1\Gyr$,
the maximum expected dust mass is $\sim 10^{-4} M_\oplus$ at $r_{\rm d} = 10\AU$
and $\sim 10^{-5} M_\oplus$ at $r_{\rm d} = 6\AU$.
Comparing with Tab.~\ref{tab:results}, we conclude that the amount of dust in our systems 
is close to, or even somewhat greater than, the theoretical maximum 
allowed by a steady-state collisional cascade.
This means that we might have
a similar difficulty that exists in explaining other 
systems with hot excesses that have been known before, such as HD~69830
\citep{beichman-et-al-2005b}.
Proposed scenarios for such systems include:
supply of comets from an outer massive cometary reservoir, possibly
following a recent dynamical instability such as the Late Heavy Bombardment;
the inward-scattering and desintegration of a large object from such an outer
reservoir;
a recent major collision between two large planetesimals
\citep[see][and references therein]{payne-et-al-2009}.
Finally, a possibility of a steady-state collisional dust production
can be resuscitated if one allows the asteroids in the belt to have very eccentric 
orbits \citep{wyatt-et-al-2010}.
Such a belt could result from shepherding
and scattering of an initial planetesimal belt during the inward
migration of ``hot Jupiters''
\citep{payne-et-al-2009}.

\section*{Acknowledgments}

We are most grateful to the anonymous reviewer for valuable and constructive comments
that helped to improve the manuscript significantly.
Our thanks go to Ronny Errmann for his assistance in calculating the interstellar
extinction.
This research has made use of the Exoplanet Orbit Database
and the Exoplanet Data Explorer at \url{exoplanets.org}.
Our work was funded by the German DFG,
grants Kr~2164/9-1 and Lo~1715/1-1.
SF acknowledges support of the State of Thuringia via the graduate student
fellowship.


\newcommand{\AAp}      {A\&A}
\newcommand{\AApSS}    {AApSS}
\newcommand{\AcA}      {AcA}
\newcommand{\AdvSR}    {Adv. Space Res.}
\newcommand{\AJ}       {AJ}
\newcommand{\AN}       {AN}
\newcommand{\ApJ}      {ApJ}
\newcommand{\ApJL}      {ApJL}
\newcommand{\ApJS}     {ApJS}
\newcommand{\ApSS}     {Ap\&SS}
\newcommand{\ARAA}     {ARA\&A}
\newcommand{\BAAS}     {BAAS}
\newcommand{\CelMech}  {CM\&DA}
\newcommand{\EMP}      {EMP}
\newcommand{\EPS}      {EPS}
\newcommand{\JGR}      {JGR}
\newcommand{\MNRAS}    {MNRAS}
\newcommand{\PASJ}     {PASJ}
\newcommand{\PASP}     {PASP}
\newcommand{\PSS}      {PSS}
\newcommand{\RAA}      {RAA}
\newcommand{\Science}  {Sci}

\label{lastpage}


\end{document}